# Scattering in crystals with effective radiation length in the micron range

V.M. Biryukov◆

*Institute for High Energy Physics, Protvino, 142281, Russia*

**Abstract**

In computer simulations we find that particles can scatter in bent crystal lattices with effective radiation length down to a few micron or 300-700 times shorter than in the corresponding amorphous materials. We derive a theoretical estimate independent of energy for the effective radiation length that is in agreement with our simulations for C, Si, Ge, and W crystal lattices. We show that in the ongoing collimation experiment at the Tevatron a crystal-based "smart scattering material" could outperform a channeling crystal in efficiency of collimation reducing the local background rate by a factor of 40.

**PACS codes:**

**61.85.+p Channeling phenomena**

## 1. Introduction

Crystal lattice can trap and channel particle beam along a major crystallographic direction [1]. In a bent crystal, the channeled particles follow the bend [2]. This made a basis for an elegant technique of beam steering by bent channeling crystals experimentally demonstrated from 3 MeV [3] to 1 TeV [4]. In IHEP, crystal systems extract protons on permanent basis from 70 GeV ring with 85% efficiency [5]. Bent crystals channel in good agreement with predictions up to the highest energies [6-8]. The nonchanneled particles in a bent crystal are often treated as "random" particles in a misaligned lattice, i.e. they scatter on individual atoms like in amorphous body.

Recent experiments on beam collimation using bent crystals in the rings of Relativistic Heavy Ion Collider [7] and Tevatron [8] revealed strong coherent effects observed in a broad angular range. The scattering of nonchanneled particles in a bent crystal differed strongly from the scattering in amorphous body over the whole arc of the crystal bend. The same effect was observed in simulations in quantitative agreement [7,8] with RHIC and Tevatron data.

Simulations found two factors responsible for the effect [7,9]. Within the arc of the crystal bend, a particle can become tangential to atomic planes somewhere in the crystal bulk. Two effects in these conditions are known from the physics of channeling and quasi-channeling: "volume capture" (scattering-induced transfers of random particles to channeled states leading to deflection towards the atomic plane bending) [10] and "volume reflection" (scattering of random particles off the potential of bent atomic planes leading to deflection opposite the atomic plane bending) [11,12]. Volume capture happens with a low probability (≤ 1% at high-GeV energy) but may cause a big deflection, tens of μrad and much more. Volume reflection takes place with probability of ~100% but causes a deflection on the order of 1-2 critical angle (6.8 μrad at 1 TeV for Si(110) potential well of 22.8 eV). Both effects were studied experimentally on external beams from 3 MeV to high-GeV range [12].

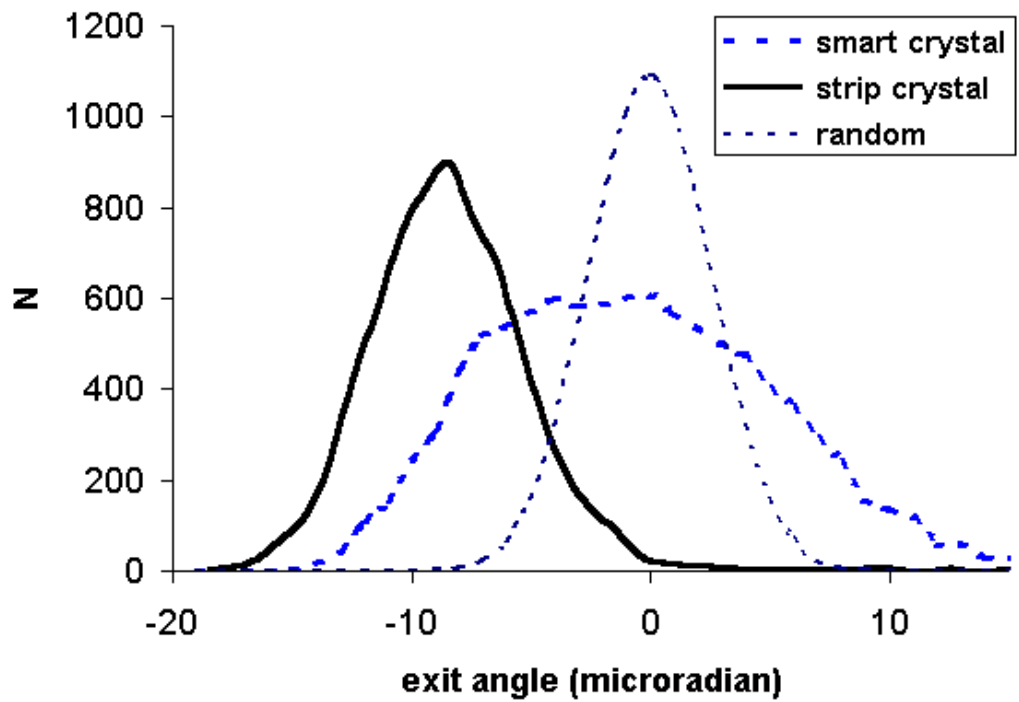

**Figure 1** The 980-GeV proton angular distribution downstream of a 3-mm Si(110) crystal: reflected from the "strip" crystal, scattered in a "smart" crystal, and at random alignment.

On a circulating beam in accelerator ring, the importance of both volume capture and reflection amplifies strongly. As requested by theory [13], the typical size of a Si crystal in modern experiments on crystal-based collimation and extraction is just 1-5 mm, or ~0.2-1% of proton nuclear interaction length [5,7]. Similar crystal size is proposed for the LHC collimation [14]. Therefore, each circulating particle may encounter a crystal up to ~100-500 times. This greatly increases the overall probability of volume capture and the overall effect of multiple volume reflections. There is another essential consideration. In external beams, bent crystals deal with deflections on the order ~30 mrad, while crystal collimation in a TeV-range ring involves a deflection of mere ~30 μrad. Such a tiny deflection requires a channelling in crystal over a distance of just ~1/1000 of the usual. This makes the probability of such an event (caused by volume capture) stronger by a huge factor. Both volume capture and reflection were found important for explanation of RHIC and Tevatron observations [7,9].

## 2. Simulations and analytical theory

Fig. 1 shows a reflection effect predicted for 980-GeV proton in a "strip" crystal (3-mm Si(110) bent 0.15 mrad) now installed into vacuum chamber of the Tevatron ring [8]. The reflected beam is shifted in angle by –8.5 μrad compared to the beam scattered in the same crystal at random alignment (shown). A barely seen tail of volume-captured protons on the right of the reflected peak amounts to ~1% protons for deflections greater than 30 μrad. Such a nice reflection happens only for crystal curvature $1/R$ much less than a critical one [11], which is $pv/R$ ≈6 GeV/cm in Si for beam of momentum $p$ and velocity $v$. In general, particle distribution shows both a shift in the mean angle and an increase in rms angle because of coherent scattering in the field of bent atomic planes. Fig. 1 shows another example with the same crystal bent to the curvature of 4 GeV/cm: the mean angle is 0 but the rms angle is greater than one at random alignment. We marked it a "smart" crystal because in the following we explore bent crystals as a smart material, which exhibits unique scattering properties within a given angular range (the arc of the crystal bend) but shows the usual, amorphous-like scattering behavior outside of this angular range.

Fig. 2 shows the rms and mean exit angles (in terms of the critical angle) of 1-TeV proton downstream of a 1-mm bent Si(110) crystal obtained in simulations as a function of the crystal curvature (in terms of the critical curvature). For a strong curvature, the mean angle becomes zero while the rms angle is still increased compared to the rms scattering angle in a misaligned crystal (1.45 μrad or 0.2 critical angle).

A particle circulating in accelerator ring may scatter in a bent crystal many times. Between the scatterings, particle makes a number of turns (and many betatron oscillations) in the ring and comes to the crystal with some new incidence angle. If the new angle is still within the arc of the crystal bend, the particle will experience another coherent scattering, and so on. The rms angle $\theta_{rms}$ in individual scatterings is independent of the incidence angle if it is within the arc of the crystal bend. Particle scattering with rms angle $\theta_{rms}$ over a crystal length $L$ can be characterized by effective radiation length $L_{eff}$ defined by analogy to radiation length $L_R$ of multiple Coulomb scattering:

$$\theta_{rms} \approx \frac{E_S^2}{p^2 v^2} \frac{L}{L_{eff}} \qquad (1)$$

with $E_S \approx 14$ MeV. Simulations showed that in crystal collimation experiments [7,8] with a 5-mm Si(110) crystal the effective radiation length $L_{eff}$ was shorter than the usual $L_R = 9.38$ cm in amorphous Si by a factor of 3 in RHIC case and factor of 5 in Tevatron case, which led to significant and even dramatic changes measured in particle loss rate when the crystal alignment was within the angular range corresponding to its arc bend ≈0.44 mrad. The increase in crystal scattering angle from coherent effects made a beam diffusion to increase several-fold.

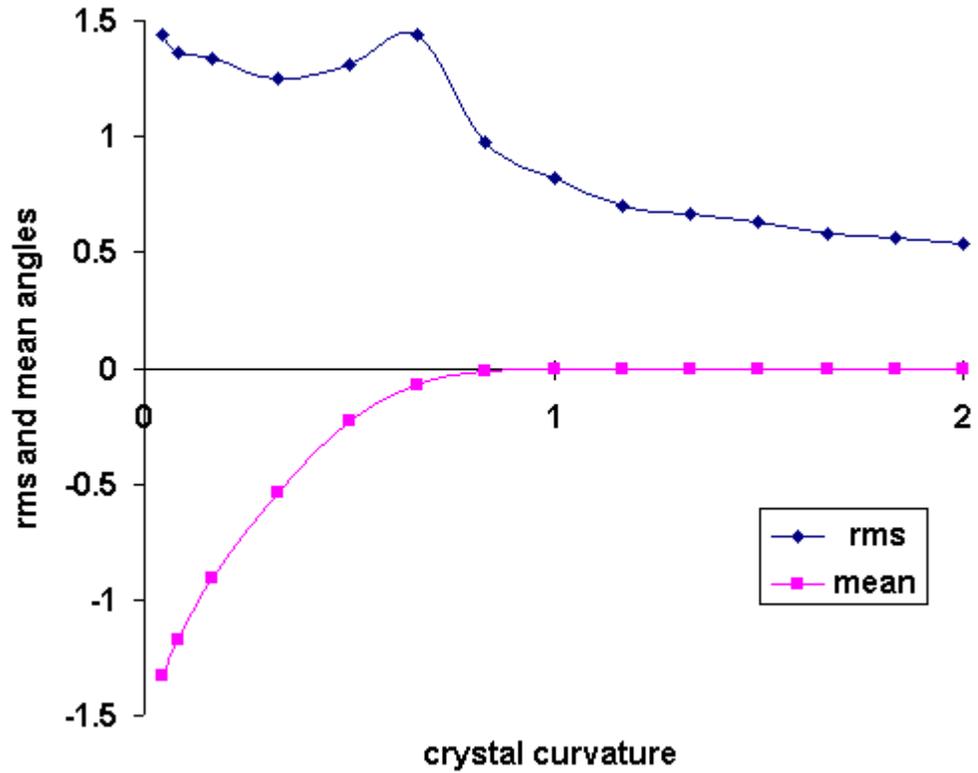

**Figure 2** The rms and mean exit angles (in terms of the critical angle) of proton downstream of a bent Si(110) crystal as a function of the crystal curvature (in terms of the critical curvature).

We studied in computer simulations a 7-TeV proton scattering in several crystal lattices of different atomic number $Z$: C(110), Si(110), Ge(110), and W(110), trying to find in every case the limits of the effective radiation length $L_{eff}$. The results can be better understood after a simple analytical consideration.

We can show that for a particle scattering in crystal lattice bent with a given angle $\theta$ the minimal effective radiation length $L_{eff}$ is independent of energy. On this way, we also derive a simple analytical estimate for a minimal possible $L_{eff}$ in a crystal lattice. The minimal curvature radius when the scattering is strong is on the order of the critical radius $R_C$. With the crystal bend angle of $\theta$, the crystal length is $L \approx R_C \theta$. The angle of coherent scattering over this distance is on the order of $\theta_C$. We start with equation

$$\theta_C^2 \approx \frac{E_S^2}{p^2 v^2} \frac{R_C \theta}{L_{eff}} \qquad (2)$$

and estimate the effective radiation length $L_{eff}$ as

$$L_{eff} \approx \frac{E_S^2}{p^2 v^2} \frac{R_C \theta}{\theta_C^2} \qquad (3)$$

from which one can already see that it is independent of energy because $R_C \sim pv$ and $\theta_C \sim (pv)^{-1/2}$.

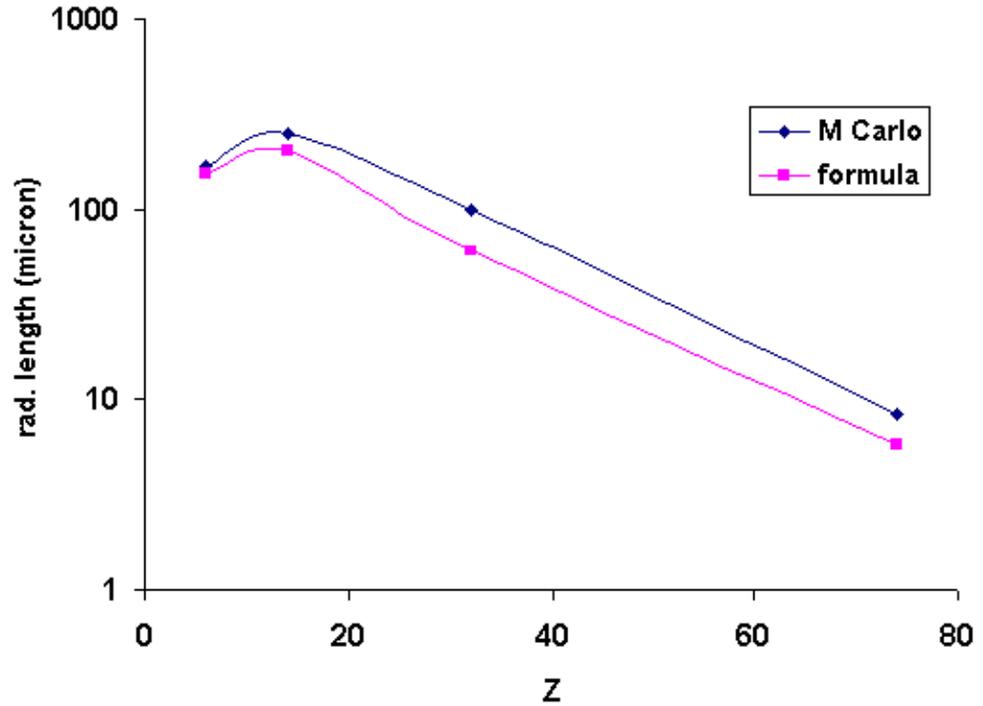

**Figure 3** Effective radiation length as computed by Eq. (4) and as obtained in simulations for several crystal lattices: C(110), Si(110), Ge(110), and W(110) shown as a function of a crystal atomic number $Z$.

We take the constant $E_S^2 \approx 4\pi(m_e c^2)^2/\alpha$ in definition Rossi-Greisen [15], approximate $R_C$ and $\theta_C$ with Lindhard potential as in ref. [16] and obtain a very simple estimate:

$$L_{eff} \approx \frac{137\,\theta}{\sqrt{3\pi a_B r_e^2 N^2 d_p^2 Z^{\frac{5}{3}}}} \qquad (4)$$

Here 137 is inversed α ≈1/137, $a_B$ Bohr radius, $r_e$ classical electron radius, $N$ volume density of atoms, $d_p$ interplanar distance, $Z$ atomic number of crystal. The angle $\theta$, which sets the angular range where $L_{eff}$ is effective, becomes a control parameter that can be used to set $L_{eff}$ as required in application. The smaller is the angular range of the effect, the shorter $L_{eff}$ can be achieved.

Figure 3 shows $L_{eff}$ (for $\theta$=25 μrad) as computed by Eq. (4) and actual values as obtained in Monte Carlo simulations for crystal lattices from C(110) to W(110); eq (4) describes data rather well. These examples show $L_{eff}$ reduced by a factor of 300-700 compared to a corresponding amorphous material (or randomly aligned crystal), bringing $L_{eff}$ value below 10 μm in case of W(110). The highest reduction factor, 700, is achieved for diamond. The value of $\theta$=25 μrad chosen in the examples is practical in applications like beam collimation in the LHC ring.

Fig. 4 shows an example of 7 TeV proton scattering in a Si(110) crystal bent 30 μrad in two cases: under coherent scattering conditions (smart properties switched on, effective within the range 30 μrad wide) and at random alignment (i.e. smart properties switched off). Notice that with planar orientation of crystal lattice the effect of strong coherent scattering takes place in just one plane (the plane of bending) while in the normal plane the scattering remains the usual amorphous-like.

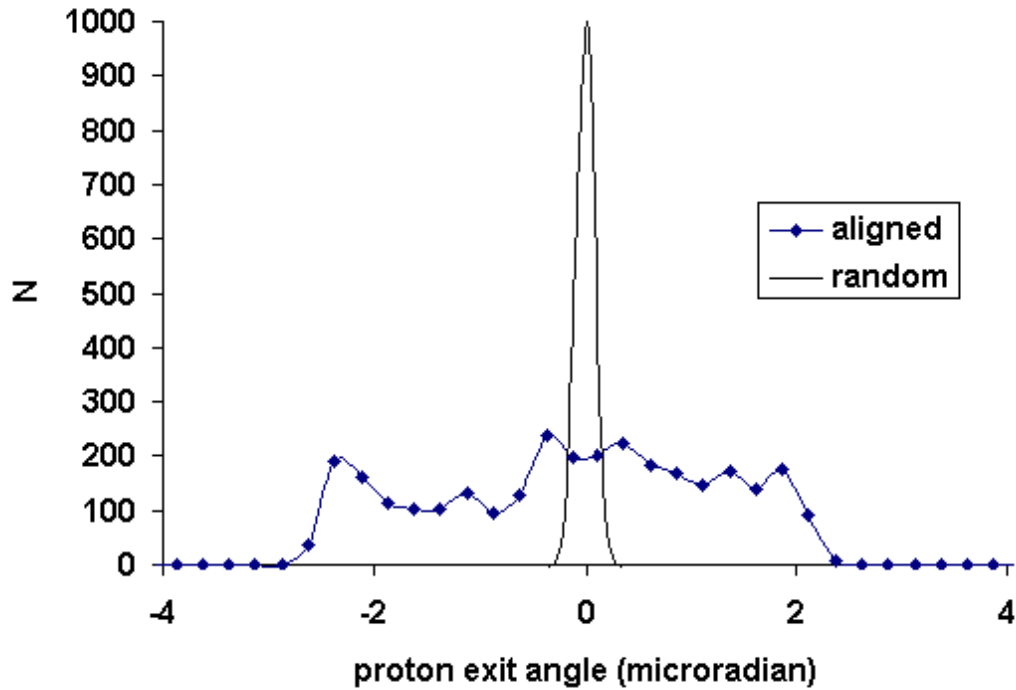

**Figure 4** Exit angular distributions for 7 TeV proton in a Si(110) crystal under coherent scattering conditions and at random alignment.

The simulations have also shown that the nuclear interaction length $L_N$ in a bent crystal lattice is almost the same as one in amorphous material. According to simulations, the $L_N/L_{eff}$ ratio can run up to 12000 while in amorphous materials $L_N/L_R$ is order of 5 to 30. The value of $L_N$ scales with atomic weight $A$ roughly as $A^{-2/3}$, so we expect the $L_N/L_{eff}$ ratio to be scaled with $Z$ as

$$\frac{L_N}{L_{eff}} \sim \frac{1}{\theta} Z \left(\frac{Z}{A}\right)^{\frac{2}{3}} \qquad (5)$$

Recalling a formula for radiation length $L_R$ in amorphous material, we expect the rough scaling of the $L_R/L_{eff}$ ratio with $Z$ as

$$\frac{L_R}{L_{eff}} \sim \frac{1}{\theta} Z^{-\frac{1}{3}} \qquad (6)$$

In all formulae (3)-(6) we have $\theta$ as control parameter that can be used for tuning of the "smart material" properties. Notice also that the unique properties are effective within the angular range $\theta$ only. By tilting the sample outside this angular range, one switches the smart material properties back to normal amorphous values.

The above consideration doesn't mean that a bending angle $\theta$ can be realized as small as one wants in any energy range. Actually, the minimal crystal size should be on the order of one oscillation length $\lambda$ in a planar channel and the bending radius cannot be much smaller than $R_C$. Therefore, the angular range $\theta$ where a smart material is effective should be greater than $\sim\lambda/R_C$, with respective restriction on the achievable (3)-(6) values. The $\lambda/R_C$ value is about 20 µrad at 7 TeV in silicon, which is quite practical figure for applications in LHC collimation. With energy $E$ the $\lambda/R_C$ value is scaled as $E^{-1/2}$ and becomes ~20 mrad in low MeV range.

## 3. Tevatron test of "smart scattering material"

The unique scattering properties can be observed in just a single interaction of particle with a bent crystal lattice, as shown in Figs. 1 and 4. However, the accelerator ring with a circulating beam offers spectacular opportunities for the studies of crystal-based smart materials, as already demonstrated at RHIC [7] and Tevatron [8]. The ongoing Tevatron experiment on crystal collimation is an excellent test bed for new crystal smart materials.

It can measure nuclear interaction rate in a crystal irradiated by a circulating beam of 980 GeV protons over a broad range of crystal alignment. Fig. 5 shows the predicted dependence of the rate on alignment for the new, "strip-type" crystal (3-mm Si(110) bent 0.15 mrad) now installed into vacuum chamber of the Tevatron ring [8].

For simulations we used Monte Carlo code CATCH [17] with the recent edition of Tevatron accelerator lattice [18]. More details of the used settings are in ref. [19]. The crystal was placed at 5σ and served as a primary element in collimation scheme. The secondary collimator was placed 31.5 m downstream, at 5.5σ. Particle tracking in the Tevatron lattice was done with linear transfer matrices. Each particle was allowed to make an unlimited number of turns in the ring and of encounters with the crystal until a particle either undergoes a nuclear interaction in the crystal or hits the secondary collimator (because of a bending effect in the channeling crystal or of the scattering events).

For best alignment of the strip crystal, one has a remarkable dip in the rate, 96% down from the rate observed at random orientation, due to channeling of protons with high efficiency onto the secondary collimator. The plateau 0.15 mrad wide (the arc of the strip crystal) is the effect of coherent scattering where the effective radiation length is 7 mm, factor of 13 shorter than in amorphous Si, and the rate is suppressed by 65%. The particle amplitude in accelerator grows faster if coherent processes contribute strongly to the overall scattering. The difference in beam dynamics on the phase space with a crystal under conditions of coherent scattering leads to a faster particle loss on the secondary elements of accelerator. As a result, the particle loss (nuclear interactions) in crystal is reduced, as part of the loss now goes to different elements in the ring.

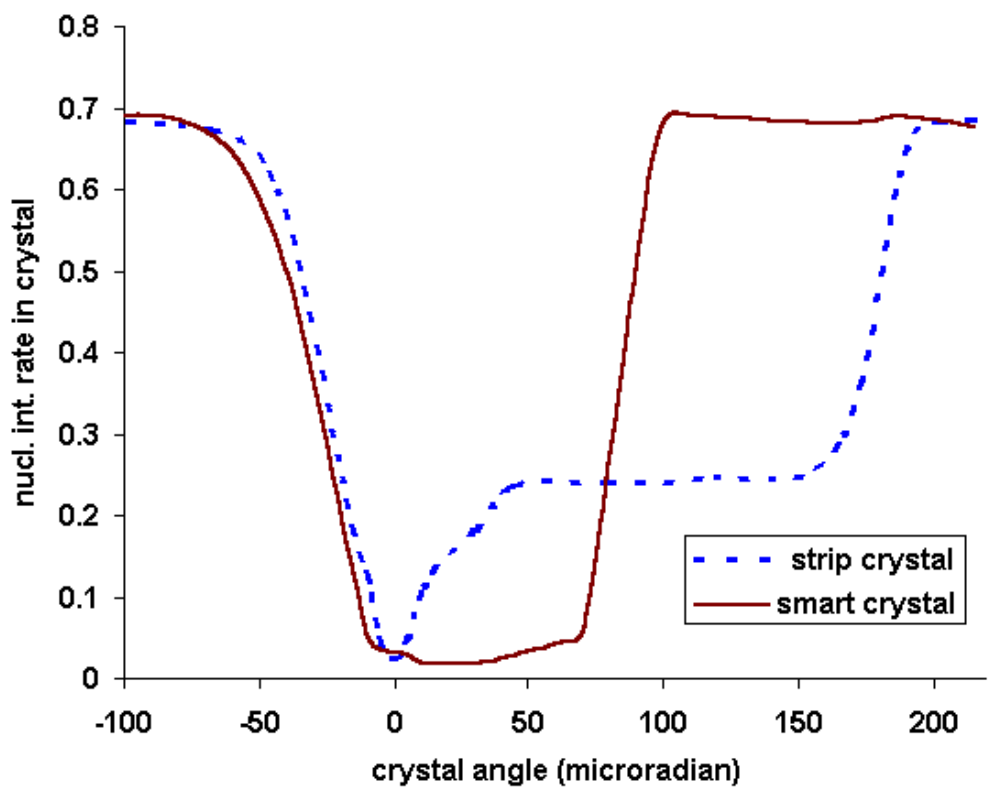

**Figure 5** The predicted crystal nuclear interaction rate for protons in the Tevatron with the "strip" crystal and with a "smart" crystal.

In order to check our understanding of crystal-based smart scattering materials and show their capability, we make a further prediction. In the same collimation setup we simulated a Si(110) crystal 0.1 mm along the beam with bending of 60 μrad. Notice that there are techniques producing Si crystals as short as 1 μm -1 mm along the beam, with cross size in centimeter range and bending up to mrad range, which can be very useful for considered test and applications

[12]. Our choice for a smart crystal was the curvature of 6 GeV/cm, which excludes any possibility of channeling. Fig. 5 shows the prediction for the simulated smart crystal in comparison with a channeling strip crystal.

The expected effective radiation length within the arc of 60 μrad is only 0.5 mm or factor of 180 (!) shorter than its value outside of the arc (or in amorphous Si). It produces a tremendous effect bringing the rate down by 97.5% or factor of 40 from the random rate. What is particularly striking is that the effect is stronger than even channeling (which itself shows a record efficiency here). In the channeling crystal, the rate within 5 μrad of the tip is 0.075 of the random rate whereas in the "smart" crystal the rate is 0.025, i.e. three times lower than in the channeling crystal. Moreover, smart crystal shows the low rate over a broad angular range. In accelerator ring, such a smart target would be an extremely efficient scatterer.

## 4. Conclusion

We conclude that coherent effects in particle scattering in bent crystal lattices can reduce the effective scattering length by 2-3 orders of magnitude compared to radiation length in amorphous materials. This phenomenon was behind the effects observed in crystal collimation at RHIC and Tevatron where the effective radiation length was shorter by a factor of 3 to 5 than in amorphous silicon, according to simulations. Bent crystals can be realized as smart materials showing the effective radiation length in a single plane 300-700 times shorter within a defined angular range, with normal radiation length of amorphous material outside of that angular range. A simple theoretical estimate is found for the effective radiation length that is in agreement with our simulations for C, Si, Ge, and W crystal lattices. In crystal collimation experiment, the new smart material can outperform the channeling crystal by a big factor, as predictions for the ongoing Tevatron experiment show. More possibilities may come from nanostructures where beam reflection is also observed in simulations [20]. This could make a basis for a new principle of beam instrumentation in accelerators and improve the efficiency of radiation protection by orders of magnitude, e.g. in collimation systems of LHC and ILC.